\documentclass[a4paper]{JHEP3}

\usepackage[english]{babel}
\usepackage{amsmath,amssymb}
\usepackage[dvips]{graphicx}
\usepackage{ifthen,array}
\usepackage[comma,square,sort&compress]{natbib}
\usepackage{amsfonts}
\usepackage{bbm}

\allowdisplaybreaks

\newcommand{\flux}[2][]{\ensuremath{\ifthenelse{\equal{#1}{}}{}{^{#1}\!}
\mathit{#2}}}

\newcommand{\AddrAHEP}{
  AHEP Group, Instituto de F\'{\i}sica Corpuscular --
  C.S.I.C./Universitat de Val{\`e}ncia \\
  Edificio Institutos de Paterna, Apt 22085, E--46071 Val{\`e}ncia, Spain}

 \title{A simple analytic three--flavour description of
  the day-night effect in the solar neutrino flux }

\author{ E. Kh. Akhmedov\thanks{On leave from the National Research
    Centre Kurchatov Institute, Moscow, Russia}~, M.~A.~T{\'o}rtola and 
  J.~W.~F.~Valle \\
  \AddrAHEP \\
  E-mail:  \email{akhmedov@ific.uv.es}, \email{mariam@ific.uv.es},
  \email{valle@ific.uv.es}}

\abstract{
  In the 3-flavour framework we derive a simple approximate analytic
  expression for the day-night difference of the flux of solar $\nu_e$
  at terrestrial detectors which is valid for an arbitrary Earth
  density profile. Our formula has the accuracy of a few per cent and
  reproduces all the known analytic expressions for the Earth matter
  effects on the solar neutrino oscillations obtained under
  simplifying assumptions about the Earth's density profile (matter of
  constant density, 3 layers of constant densities, and adiabatic
  approximation).  It can also be used for studying the Earth matter 
  effects on the oscillations of supernova neutrinos.  We also discuss 
  the possibility of probing the leptonic mixing angle $\theta_{13}$ 
  through day--night asymmetry measurements at future water Cherenkov 
  solar neutrino detectors. We show that, depending on the measured value 
  of the asymmetry, the current upper bound on $\theta_{13}$ may be 
  improved, or even a lower bound on this mixing parameter may be 
  obtained.
  }


\preprint{IFIC/04-13 \\
  hep-ph/0404083}

\begin{document}

\section{Introduction}

Solar neutrinos coming to terrestrial detectors at night travel some
distances inside the Earth and so their oscillations are affected by
the Earth's matter. This leads to a difference between the day-time
and night-time solar neutrino signals -- the day-night, or
``regeneration'' effect
\cite{MS,Bouchez:1986kb,Cribier:1986ak,Carlson:1986ui,Baltz:1987hn,Dar:1987pj,
Gonzalez-Garcia:2000dj}.
Solar neutrino oscillations~\footnote{The presence of non-oscillation
phenomena is now strongly constrained by a combination of KamLAND and
solar neutrino data, see, e.~g.  Ref.~\cite{pakvasa:2003zv} and
references therein.} depend mainly on two parameters, the mixing angle
$\theta_{12}$ and the mass squared difference $\Delta m_{21}^2$. For
example, the recent three--neutrino global fit of solar, atmospheric,
reactor and accelerator neutrino data of Ref.~\cite{Maltoni:2003da}
gives for the solar neutrino oscillation parameters the $3\sigma$
allowed ranges $\theta_{12}=(28.6\div 38.6)^\circ$ and $\Delta
m_{21}^2=(5.4\div 9.5)\times 10^{-5}$ eV$^2$, with the best-fit values
$\theta_{12}=33.2^\circ$ and $\Delta m_{21}^2=6.9\times 10^{-5}$
eV$^2$.  In the three-flavour framework there is an additional
dependence on the mixing angle $\theta_{13}$ which determines the
component of the electron neutrino in the third mass eigenstate
$\nu_3$ separated by a large mass gap from the other two mass
eigenstates. This mixing parameter is mainly constrained by the data
of the CHOOZ reactor experiment \cite{apollonio:1999ae} and is known
to be small,
\begin{equation}
\theta_{13} \lesssim 9.8^\circ~(13.4)^\circ\,, \quad\mbox{or}\quad
\sin \theta_{13} \lesssim 0.17~(0.23)
\label{eq:th13bound}
\end{equation}
at 90\% C.L. (3$\sigma$) \cite{Maltoni:2003da}. The smallness of
$\theta_{13}$ implies that its effect on the solar neutrino
oscillations is rather mild; however, with increasing accuracy of the
data even relatively small effects have to be taken into account.

Three-flavour analyses of the solar neutrino data, including the 
day-night effect, have been performed in a number of recent studies
\cite{Maltoni:2003da,Bandyopadhyay:2003pk,deHolanda:2003nj,
Gonzalez-Garcia:2003qf,Fogli:2002au,Blennow:2003xw}.
In particular, in Ref.~\cite{Blennow:2003xw} a simplified analytic
treatment of the day-night effect in the three--flavour framework was
given, and it was shown that the day-night $\nu_e$ flux difference
scales essentially as $\cos^6\theta_{13}$ rather than
$\cos^4\theta_{13}$, which is the expected scaling law for the
day-time flux.  The authors of \cite{Blennow:2003xw} considered the
Earth matter effect on the solar neutrino oscillations under the 
assumption of constant matter density of the Earth. The
purpose of the present paper is to extend their analysis beyond the
constant density approximation, and to scrutinize to what extent
day--night asymmetry measurements may become a significant way to
probe for the mixing angle $\theta_{13}$. Using the relative smallness
of the matter-induced potential of neutrinos inside the Earth, we
derive a simple and accurate analytic formula for the day-night flux
difference in the case of an arbitrary Earth's density profile. We
compare our result with the known analytic formulas obtained under
simplifying assumptions about the density profile of the Earth as well
as with the results of the exact numerical calculations.

The paper is organized as follows. In sec. \ref{sec:day-night-effect}
a general consideration of the day-night effect in the 3-flavour
framework is given. In sec. \ref{sec:an-analyt-expr} we present our
main result -- the derivation of an approximate 3-flavour analytic
formula for the day-night $\nu_e$ flux difference in the case of an
arbitrary Earth density profile.  We also discuss a number of special
cases for which the day-night flux difference can be found
analytically.  These include approximating the matter density profile
of the Earth by one layer or three layers of constant densities, and
also the adiabatic approximation to neutrino evolution inside the
Earth.  In sec.  \ref{sec:comp-with-numer} the comparison of our
formulas with the results of the exact numerical calculations is
performed. In sec.~\ref{sec:future} we make a quantitative study of
the simple correlation between the day--night asymmetry in the solar
neutrino flux and the magnitude of the mixing angle $\theta_{13}$
characterizing leptonic CP violation in neutrino
oscillations~\footnote{There are additional CP violating phases but
  they do not affect neutrino oscillations, only lepton-number
  changing processes, like neutrino-less double beta
  decay~\cite{schechter:1980gr,schechter:1981gk,bilenky:1980cx,doi:1981yb}.}.
We use this correlation in order to forecast the degree to which the
mixing angle $\theta_{13}$ can be probed at future solar neutrino
measurements of the day--night asymmetry. We discuss our results in
sec.~\ref{sec:disc-outlook}. In the Appendix, we present the derivation of 
an improved perturbation-theoretic formula for the Earth regeneration factor.

\section{Day-night effect with three flavours: Generalities}
\label{sec:day-night-effect}

The 3-flavour \textsl{day-time} probability of finding an electron
neutrino produced in the Sun in the electron neutrino state at the
detector is to a very good accuracy given by \cite{Lim}
\begin{equation}
P_D=c_{13}^4 P_{2f}(\theta_{12}, \Delta m_{21}^2, c_{13}^2 V) + 
s_{13}^4\,.
\label{PD}
\end{equation}
Here $P_{2f}(\theta_{12}, \Delta m_{21}^2, c_{13}^2 V)$ is the
2-flavour $\nu_e$ survival probability in the Sun, calculated with the
neutrino potential in the Sun $V$ substituted by the effective
potential $c_{13}^2 V$, and we use the standard notation
$s_{ij}=\sin\theta_{ij}$, $c_{ij}=\cos\theta_{ij}$. Upon averaging
over fast neutrino oscillations inside the Sun, this probability can
be written as
\begin{equation}
P_{2f}(\theta_{12}, \Delta m_{21}^2, c_{13}^2 V)=\frac{1}{2}\left[1+\cos 
2\theta_{12}\, \overline{\cos 2\hat{\theta}_{12}}(1-2 P')\right]\,,
\label{P2f}
\end{equation}
where $P'$ is the probability of transition between the first and
second matter eigenstates in the course of the neutrino propagation in
the Sun (``hopping'' probability), $\hat{\theta}_{12}(r)$ is the
effective mixing angle in matter at the point $r$, and $\overline{\cos
  2\hat{\theta}_{12}}$ is the value of $\cos 2\hat{\theta}_{12}$
averaged over the neutrino production point \cite{Blennow:2003xw}:
\begin{equation}
\overline{\cos 2\hat{\theta}_{12}}=\int_0^{R_\odot} \!dr f(r) \cos 
2\hat{\theta}_{12}(r) \,.
\end{equation}
Here $f(r)$ is the normalized spatial distribution function of the
neutrino source in the Sun~\cite{bahcall}, and the explicit formula
for $\cos 2\hat{\theta}_{12}(r)$ will be given below [Eqs.
(\ref{cos}) and (\ref{omega})].

The solar neutrino flux arriving at the Earth is an incoherent sum of
the fluxes of mass-eigenstate neutrinos $\nu_1$, $\nu_2$ and $\nu_3$
(see, e.g., \cite{Dighe:1999id}); therefore the day-time $\nu_e$ 
survival probability can in general be written as 
\begin{equation}
P_D=P_{e1}^\odot\,P_{1e}^{(0)}+P_{e2}^\odot\,P_{2e}^{(0)}+P_{e3}^\odot\,
P_{3e}^{(0)}\,,
\label{PD1}
\end{equation}
where $P_{ei}^\odot$ ($i=1,2,3$) is the probability of $\nu_e\to
\nu_i$ conversion in the Sun, and $P_{ie}^{(0)}$ is the projection of
the $i$th mass eigenstate onto $\nu_e$: $P_{ie}^{(0)}=|U_{ei}|^2$, $U$
being the leptonic mixing matrix in vacuum. For this matrix we use the
standard parameterization \cite{hagiwara:2002fs}
\begin{align}
U &=  O_{23} \Gamma_\delta O_{13} \Gamma^\dagger_\delta O_{12} \nonumber\\
  &= \left( \begin{matrix} c_{12} c_{13} &
    s_{12} c_{13} & s_{13} {\rm e}^{-{\rm i} \delta_{\rm CP}} \\ -s_{12}
    c_{23} - c_{12} s_{13} s_{23} {\rm e}^{{\rm i} \delta_{\rm CP}} & c_{12}
    c_{23} - s_{12} s_{13} s_{23} {\rm e}^{{\rm i} \delta_{\rm CP}} & c_{13}
    s_{23} \\ s_{12} s_{23} - c_{12} s_{13} c_{23} {\rm e}^{{\rm i}
      \delta_{\rm CP}} & -c_{12} s_{23} - s_{12} s_{13} c_{23} {\rm
      e}^{{\rm i} \delta_{\rm CP}} & c_{13} c_{23} \end{matrix} \right)\,,
   \label{eq:U}
\end{align}
where $O_{ij}$ is the orthogonal rotation matrix in the $ij$-plane
which depends on the mixing angle $\theta_{ij}$, and
$\Gamma_\delta=\mathrm{diag} (1,1,{\rm e}^{{\rm i}\delta_{\rm CP}})$,
$\delta_{\rm CP}$ being the Dirac-type CP-violating
phase. 

As we shall see shortly (see also the discussion in Ref.~\cite{Blennow:2003xw}), 
the third matter eigenstate $\nu_{3m}$ inside both the Sun and the Earth 
essentially decouples from $\nu_{1m}$ and $\nu_{2m}$. In addition, it is 
practically not affected by the solar or Earth matter, i.e. $\nu_{3m} \simeq 
\nu_{3}$. Therefore one has $P_{e3}^{\odot}\simeq P_{3e}^{(0)}=s_{13}^2$. It 
is also easy to show that upon averaging over the fast neutrino oscillations 
in the Sun one obtains \cite{Blennow:2003xw}
\begin{equation}
P_{e1}^\odot=\frac{c_{13}^2}{2}\left[1+\overline{\cos 2\hat{\theta}_{12}}(1-
2 P')\right]\,,\qquad\quad P_{e2}^\odot=\frac{c_{13}^2}{2}\left[1-
\overline{\cos 2\hat{\theta}_{12}}(1-2 P')\right]\,.
\label{P1P2}
\end{equation}
Substituting (\ref{P1P2}) into (\ref{PD1}), one recovers Eq.~(\ref{PD}) with 
$P_{2f}(\theta_{12}, \Delta m_{21}^2, c_{13}^2 V)$ given by Eq.~(\ref{P2f}). 
Note that for the LMA MSW solution of the solar neutrino problem $P'$ is 
extremely small; therefore we shall neglect it from now on. 

Since the flux of solar neutrinos coming to the Earth is an incoherent sum of 
the fluxes of mass-eigenstate neutrinos, the \textsl{night-time} $\nu_e$ 
survival probability at the detector can be written as 
\begin{equation}
P_N=P_{e1}^\odot\,P_{1e}^{\oplus}+P_{e2}^\odot\,P_{2e}^{\oplus}+P_{e3}^\odot\,
P_{3e}^{\oplus}\,,
\label{PN1}
\end{equation}
where $P_{ie}^\oplus$ is the probability that a neutrino arriving at
the Earth as a mass eigenstate $\nu_i$ is found at the detector in the
$\nu_e$ state after having traveled a distance $L$ inside the Earth.
The night-day probability difference is therefore
\begin{equation}
P_N-P_D=P_{e1}^\odot\,
(P_{1e}^{\oplus}-P_{1e}^{(0)})
+P_{e2}^\odot\,(P_{2e}^{\oplus}-P_{2e}^{(0)})+P_{e3}^\odot\,
(P_{3e}^{\oplus}-P_{3e}^{(0)})\,.
\label{PND1}
\end{equation}
The probabilities $P_{ei}^\odot$, $P_{ie}^\oplus$ and $P_{ie}^{(0)}$ satisfy  
the conditions 
\begin{equation}
\sum_{i=1}^3 P_{ei}^{\odot}~=~\sum_{i=1}^3 P_{ie}^{\oplus}~=~\sum_{i=1}^3 
P_{ie}^{(0)}~=~1\,,
\label{unit}
\end{equation}
which follow from the unitarity of neutrino evolution. Since 
$P_{3e}^{\oplus}=P_{3e}^{(0)}$, Eq.~(\ref{unit}) gives $P_{1e}^\oplus-
P_{1e}^{(0)}=-(P_{2e}^\oplus-P_{2e}^{(0)})$. Substituting this into 
Eq.~(\ref{PND1}) yields 
\begin{equation}
P_N-P_D=(P_{e2}^\odot-P_{e1}^\odot) (P_{2e}^{\oplus}-
P_{2e}^{(0)})=-c_{13}^2\,\overline{\cos 2\hat{\theta}_{12}}\,
(P_{2e}^{\oplus}-P_{2e}^{(0)})\,,
\label{PND2}
\end{equation}
where in the last equality Eq.~(\ref{P1P2}) with $P'=0$ has been used. 
The Earth matter effects on the solar neutrino oscillations are encoded 
in the factor $P_{2e}^{\oplus}-P_{2e}^{(0)}$; we now turn 
to the calculation of this quantity. 

Neutrino evolution equation in the flavour basis can be written as 
\begin{equation}
i\frac{d}{dt}\nu=\left[U\,{\rm diag}(0,\, 2\delta,\, 2\Delta)\,U^\dagger 
+{\rm diag}(V(t),\, 0, \,0)\right]\nu \,,
\label{evol1}
\end{equation}
where $\nu=(\nu_e,\, \nu_\mu,\, \nu_\tau)^T$ is the wave function of the 
neutrino system, $t$ is the coordinate along the neutrino trajectory, and 
the parameters 
\begin{equation}
\delta =\frac{\Delta m_{21}^2}{4E}\,,\qquad\qquad
\Delta = \frac{\Delta m_{31}^2}{4E}\,
\label{deltas}
\end{equation}
involve the mass-squared differences responsible for the solar and
atmospheric neutrino oscillations, with $\Delta m_{31}^2$ large
compared to $\Delta m_{21}^2$ \cite{Maltoni:2003da}: $\Delta
m_{31}^2\sim 2\times 10^{-3}~{\rm eV}^2\sim 30\, \Delta m_{21}^2$. The
matter-induced potential of neutrinos in Eq.~(\ref{evol1}) is
\begin{equation}
V(t)=\sqrt{2}\,G_F\,N_e(t)\,,
\label{V}
\end{equation}
where $G_F$ is the Fermi constant and $N_e(t)$ is the electron number 
density in matter. 

For our purposes it proves convenient to go to the new basis 
defined through 
\begin{equation}
\nu = O_{23} \Gamma_\delta O_{13} \tilde{\nu}\,.
\label{eq:defM}
\end{equation}
The evolution equation for the rotated $\tilde{\nu}$ state is
\begin{equation}
i\frac{d}{dt}\left( \begin{array}{c}\tilde{\nu}_1 \\
\tilde{\nu}_2 \\ \tilde{\nu}_3 \end{array}\right) = 
\left( \begin{array}{ccc} 2 s_{12}^2\, \delta + c_{13}^2 V(t) & 
2 s_{12} c_{12} \,\delta & s_{13} c_{13} V(t) \\
2 s_{12} c_{12} \,\delta & 2 c_{12}^2 \,\delta & 0 \\
s_{13} c_{13} V(t) & 0 & 2 \Delta + V(t) s_{13}^2 
\end{array}\right)
\left( 
\begin{array}{c}\tilde{\nu}_1 \\
\tilde{\nu}_2 \\ \tilde{\nu}_3 \end{array}\right)\,.
\label{evol2}
\end{equation}
For neutrino evolution in the Sun and in the Earth one has $V\lesssim
2\delta \ll 2\Delta$ (actually, inside the Earth $V\ll 2\delta$).
Since, in addition, $s_{13}\ll1$, one can to a very good accuracy
neglect the (1-3) and (3-1) elements of the effective Hamiltonian in
Eq.~(\ref{evol2}) compared to the (3-3) element. This means that the
evolution of the third matter eigenstate essentially decouples from
that of the first two eigenstates and, in addition, that matter
effects on the third eigenstate are negligible. This approximation is
especially good for the $\nu_e$ survival probability since, as was
shown in \cite{Akhmedov:2001kd,Akhmedov:2004ny}, terms of first order in 
$s_{13}$ cancel in this quantity, and the correction only appears at order
$s_{13}^2 V^2/\Delta^2$.

Diagonalization of the $(\tilde{\nu}_1,\,\tilde{\nu}_2)$ - subsector of 
the effective Hamiltonian in Eq.~(\ref{evol2}) yields the instantaneous 
effective mixing angle in matter $\hat\theta_{12}(t)$, which can be 
determined through
\begin{equation}
\cos 2\hat\theta_{12}(t)=\frac{\cos 2\theta_{12}\,\delta-c_{13}^2 V(t)/2}
{\omega(t)}\,,
\label{cos}
\end{equation}
where 
\begin{equation}
\omega(t)=\sqrt{[\cos 2\theta_{12}\,\delta-c_{13}^2 V(t)/2]^2+\delta^2 
\sin^2 2\theta_{12}}\,.
\label{omega}
\end{equation}

Let us introduce the neutrino evolution matrix in the rotated basis according 
to
\begin{equation}
\tilde{\nu}(t)=\tilde{S}(t,\,t_0)
\tilde{\nu}(t_0)\,,\qquad\qquad 
\tilde{S}(t_0,\,t_0)=\mathbbm{1}\,.
\label{S}
\end{equation}
The matrix $\tilde{S}(t,\,t_0)$ satisfies the same evolution equation, 
Eq.~(\ref{evol2}), as $\tilde{\nu}$. The decoupling of the third eigenstate 
implies that it can be written as 
\begin{equation}
\tilde{S}(t,t_0)=\left(\begin{array}{ccc}
\tilde{\alpha}(t,t_0)   & \tilde{\beta}(t,t_0)    & 0 \\
-\tilde{\beta}^*(t,t_0)  &  \tilde{\alpha}^*(t,t_0) & 0 \\
0 & 0 & f(t,t_0)
\end{array}\right)\,,
\label{S1}
\end{equation}
where 
\begin{equation}
f(t,t_0)=\exp\left[{-2 i\,\Delta (t-t_0)}\right]\,,
\end{equation}
and the parameters $\tilde{\alpha}(t,t_0)$ and $\tilde{\beta}(t,t_0)$
(satisfying $|\tilde{\alpha}|^2+|\tilde{\beta}|^2=1$) are to be found
from the solution to the two-flavor problem governed by the mass
squared difference $\Delta m_{21}^2$, mixing angle $\theta_{12}$, and
matter potential $c_{13}^2 V(t)$.

It is now easy to find the probabilities $P_{1e}^{\oplus}$ and 
$P_{2e}^{\oplus}$ 
in terms of $\tilde{\alpha}$ and $\tilde{\beta}$. Direct calculation 
yields $P_{1e}^{\oplus}=c_{13}^2|c_{12}\tilde{\alpha}-
s_{12}\tilde{\beta}|^2$, ~$\,P_{2e}^{\oplus}=c_{13}^2|s_{12}\tilde{\alpha}+
c_{12}\tilde{\beta}|^2$, and 
\vspace*{0.5mm}
\begin{equation}
P_{2e}^\oplus-P_{2e}^{(0)}=c_{13}^2 [ \cos 2\theta_{12} \,|\tilde{\beta}|^2 
+ \sin 2\theta_{12} \,{\rm Re} (\tilde{\alpha}^* \tilde{\beta})]\,. 
\label{P2e1}
\end{equation}

\section{An analytic expression for $P_N-P_D$}
\label{sec:an-analyt-expr}

We shall now derive an approximate analytic expression for
$P_{2e}^\oplus-P_{2e}^{(0)}$ valid for an arbitrary matter density
profile of the Earth. Our basic point is that the neutrino potential
in the Earth $V$ is small ($V/2\delta \lesssim 0.05$), and so can be
considered as a perturbation.

We shall use the standard perturbation theory for the evolution matrix
(see, e.g. Appendix B of Ref.~\cite{Akhmedov:2004ny}). The effective
Hamiltonian in the rotated basis in Eq.~(\ref{evol2}) can be
decomposed as $\tilde{H}(t) = \tilde{H}_0 +\tilde{H}_1(t)$, where
$\tilde{H}_0$ and $\tilde{H}_1(t)$ are of zeroth and first order in
$V(t)$, respectively.  Then to first order in $V(t)$, the evolution
matrix $\tilde{S}(t,\,t_0)$ can be written as
\begin{equation}
\tilde{S}(t,\,t_0) \simeq \tilde{S}_0(t,\,t_0)-i \tilde{S}_0(t,\,t_0)
\int_{t_0}^t [\tilde{S}_0(t'\,,t_0)^{-1} \tilde{H}_1(t') \tilde{S}_0
(t'\,,t_0)]\,{\rm d}t'\,.
\label{Stilde}
\end{equation}
Both $\tilde{S}_0(t,\,t_0)$ and $\tilde{S}(t,\,t_0)$ have the form 
(\ref{S1}). For the zeroth-order matrix $\tilde{S}_0(t,\,t_0)$ we find 
\begin{equation}
\tilde{\alpha}_0=\cos \delta L+i \cos 2\theta_{12}\,\sin \delta L\,, 
\qquad  \tilde{\beta}_0 =-i \sin 2\theta_{12}\,\sin \delta L\,.
\label{ab3}
\end{equation}
Eq.~(\ref{Stilde}) then gives 
\begin{equation}
\tilde{\alpha}=\tilde{\alpha}_0 (1-iA)- i \tilde{\beta}_0 B^*\,, 
\qquad\qquad
\tilde{\beta}=\tilde{\beta}_0 (1+iA)- i \tilde{\alpha}_0 B\,, 
\label{ab4}
\end{equation}
where 
\begin{equation}
A=\frac{c_{13}^2}{2} \int_{t_0}^t \!(|\tilde{\alpha}_0|^2-
|\tilde{\beta}_0|^2)\, V dt'\,,
\qquad
B=c_{13}^2 \int_{t_0}^t \! \tilde{\alpha}_0^* \tilde{\beta}_0\, V dt'\,.
\label{AB}
\end{equation}
Substituting Eqs.~(\ref{ab4}) and (\ref{AB}) into Eq.~(\ref{P2e1}), 
one arrives, after a little algebra, at a remarkably simple result:
\begin{equation}
P_{2e}^{\oplus}-P_{2e}^{(0)}=c_{13}^4 \sin^2 2\theta_{12}\, \frac{1}{2} 
\int_0^L\! dx \,V(x) \sin[2\delta\cdot(L-x)]\,.
\label{P2e7}
\end{equation}
Here $x$ is the coordinate along the neutrino path inside the Earth and 
$L=2 R_\oplus \cos\theta_Z$, where $R_\oplus$ is the Earth's radius and 
$\theta_Z$ is the zenith angle of neutrino trajectory. The night-day flux 
difference $P_{N-D}$ is then given by (\ref{PND2}).
Equations (\ref{PND2}) and  (\ref{P2e7}) are our main results.

Strictly speaking, for perturbation theory in $V$ to be valid, two 
dimensionless parameters have to be small, namely  $V/2\delta$ and $V L$ 
(or, more precisely, $\int_0^LV dx$). While the first parameter is indeed  
very small in our case, for long enough distances travelled by neutrinos 
in the Earth the second parameter may be rather large. Nevertheless, as we 
shall see in sec.~\ref{sec:comp-with-numer}, Eq.~(\ref{P2e7}) yields very 
accurate results even for large $V L$ provided that integration over 
neutrino energies or zenith angles is involved. 
%
%
A more accurate formula can be obtained by replacing in the integrand of 
Eq.~(\ref{P2e7}) the in-vacuum oscillation phase by the corresponding 
adiabatic phase:
\begin{equation}
  \label{eq:adiab}
P_{2e}^{\oplus}-P_{2e}^{(0)}= c_{13}^4 \sin^2 2\theta_{12}\frac{1}{2} 
\int_0^L\! dx \,V(x) \sin\left[2\int\limits_x^{L}\!\omega(x')\,dx'  
\right]\,,
\end{equation}
where $\omega(x)$ is given in Eq.~(\ref{omega}). We shall show in 
sec.~\ref{sec:comp-with-numer} that this considerably improves the 
agreement with the exact results in the cases when the oscillation phase is 
large but no averaging over zenith angles or neutrino energies is performed. 
Eq.~(\ref{eq:adiab}) is based on perturbation theory in $V/2\delta$ rather 
than in $V$ and thus is valid for arbitrary values of $V L$. We give its 
derivation in the Appendix. 

It is interesting to note that Eq.~(\ref{eq:adiab}) depends on the 
adiabatic phase even though no explicit assumptions about adiabaticity 
is made in its derivation. This is related to the fact that in the 
limit $V\ll 2\delta$  the adiabaticity condition is satisfied automatically, 
irrespective of the spatial behaviour of the potential $V(x)$.  Indeed, the 
adiabaticity condition requires that the mixing angle in matter 
$\hat\theta_{12}$ change little on the length scales of the order of the 
neutrino oscillation length in matter. In the case $V\ll 2\delta$ this mixing 
angle is always close to the vacuum mixing angle $\theta_{12}$ and so changes 
little over the whole scale L which contains many oscillation lengths, 
i.e. the neutrino evolution is automatically adiabatic.

We will consider now several special cases for which
exact analytic expressions can be found for $\tilde{\alpha}$ and
$\tilde{\beta}$ (and so also for $P_{2e}^{\oplus}-P_{2e}^{(0)}$).

The matter density of the Earth changes relatively slowly both within
its mantle and core, while at the mantle-core border it jumps by about
a factor of two. Therefore as a first approximation one can consider
the matter densities in the mantle and core as being constant. In
other words, for neutrinos traversing only the mantle of the Earth one
can use the constant density approximation, whereas for core-crossing
neutrinos the Earth density can be approximated by three layers of
constant densities (mantle-core-mantle).  More accurate treatment
along these lines would include a larger number of layers, allowing
for several layers within the mantle and the core \cite{lisi:1997yc}.
 
For existing solar neutrino detectors, most of the neutrinos detected
during the night time cross only the mantle of the Earth but not its
core (the Super-Kamiokande (Super-K) detector has the largest
fractional core coverage time equal to 7\%). We therefore start with
the case of one layer of constant density.

\subsection{Matter of constant density}
In this case one readily finds
\begin{equation}
\tilde{\alpha}=\cos \omega L+i \cos 2\hat{\theta}_{12}\,\sin \omega L\,, 
\qquad \tilde{\beta}=-i \sin 2\hat{\theta}_{12}\,\sin \omega L\,,
\label{ab1}
\end{equation}
where the mixing angle in matter $\hat{\theta}_{12}$ and $\omega$ are
given by Eqs.~(\ref{cos}) and (\ref{omega}) with $V$=const.
Substituting (\ref{ab1}) into Eq.~(\ref{P2e1}) and the result into Eq.~
(\ref{PND2}), one finds
\begin{equation}
P_{2e}^\oplus-P_{2e}^{(0)}=c_{13}^4 \,\sin^2 2\theta_{12}\, \frac{V \delta}
{2 \omega^2}\,\sin^2 \omega L\,,
\label{P2e2}
\end{equation}
\begin{equation}
P_N-P_D=-c_{13}^6\,\overline{\cos 2\hat{\theta}_{12}}\,\sin^2 2\theta_{12}\, 
\frac{V \delta}{2 \omega^2}\,\sin^2 \omega L\,.
\label{PND3}
\end{equation}
This is the result obtained in Ref.~\cite{Blennow:2003xw}. In the
limit $s_{13}\to 0$ one recovers the 2-flavour day-night probability
difference in the constant-density approximation to the Earth's
density profile given in \cite{Gonzalez-Garcia:2000ve}. 
To leading order in the potential $V$ Eq.~(\ref{PND3}) simplifies to
\begin{equation}
P_N-P_D=-c_{13}^6\,\overline{\cos 2\hat{\theta}_{12}}\,\sin^2 2\theta_{12}\, 
\frac{V}{2 \delta}\,\sin^2 \delta L\,.
\label{PND4}
\end{equation}

\subsection{Three layers of constant densities}

For neutrinos crossing the core of the Earth, the Earth's density
profile can be approximated by three layers of constant densities
$\rho_1$, $\rho_2$ and $\rho_1$ and widths $L_1$, $L_2$ and $L_1$
(mantle-core-mantle). In this case one can write $\tilde{\alpha}=Z-i
W_3$, $\tilde{\beta}=-i W_1$ with real $Z$ and $W_{1,3}$, which yields
$P_{2e}^\oplus-P_{2e}^{(0)}=c_{13}^2 W_1(\cos 2\theta_{12}\,W_1+\sin
2\theta_{12}\,W_3)$. Here \cite{Akhmedov:1998ui}
\begin{eqnarray}
W_1 = 2\sin 2\theta_1 \,Y \sin \omega_1 L_1 +\sin 2\theta_2 
\sin \omega_2 L_2\,, ~~~~~\nonumber \\
W_3 = -(2\sin 2\theta_1 \,Y \cos\omega_1 L_1 +\sin 2\theta_2 
\cos\omega_2 L_2\,)\,, 
\label{W1W3}
  \end{eqnarray}
with
\begin{equation}
Y= \cos\omega_1 L_1\,\cos\omega_2 L_2-\cos (2\theta_1-2\theta_2) 
\sin\omega_1 L_1\,\sin\omega_2 L_2\,,
\end{equation} 
and $\theta_1$ and $\theta_2$ are the values of the effective mixing 
angle $\hat{\theta}_{12}$ in matter of densities $\rho_1$ and $\rho_2$, 
respectively. After a simplification one then obtains 
\begin{align}
P_{2e}^\oplus-P_{2e}^{(0)}=c_{13}^4 \,\sin^2 2\theta_{12}\, \frac{\delta}
{2 \omega_1}&(2 Y \sin \omega_1 L_1+\frac{\omega_1}{\omega_2}\sin\omega_2 
L_2) \nonumber \\
&\times \left(\frac{V_1}{\omega_1}\,2Y \sin \omega_1 
L_1+\frac{V_2}{\omega_2}\sin\omega_2 L_2\right)\,.
\label{P2e3}
\end{align}
In the 2-flavour limit $\theta_{13}\to 0$ the expression for
$P_{2e}^\oplus-P_{2e}^{(0)}$ for the 3-layer model of the density
profile of the Earth was first obtained in \cite{Minakata:1987fj}. In
the limiting cases $V_2=V_1$, $L_2=0$ or $L_1=0$ Eq.~(\ref{P2e3}) goes
into the corresponding expressions for one layer of constant density,
Eq.~(\ref{P2e2}).

To first order in $V_{1,2}\,$ Eq.~(\ref{P2e3}) simplifies to
\begin{equation}
P_{2e}^\oplus-P_{2e}^{(0)}=c_{13}^4\,\sin^2 2\theta_{12}\,\frac{1}
{2 \delta}\left[V_1 \sin \delta(2L_1+L_2)+(V_2-V_1)\sin \delta 
L_2\right] \sin \delta(2L_1+L_2)\,.
\label{P2e4}
\end{equation}

\subsection{Adiabatic approximation}
If the matter density changes slowly enough along the neutrino path,
the evolution of the neutrino system can be studied in the adiabatic
approximation.  This approximation should be especially good for
neutrinos crossing the Earth's mantle only, and can be considered as
an improvement of the constant-density approximation. For
core-crossing neutrinos one can consider the evolution inside the core
and inside the mantle as adiabatic and match the wave function of the
neutrino state at the mantle-core borders; here we discuss the
adiabatic evolution for mantle-crossing neutrinos.

In the adiabatic approximation one has \cite{Akhmedov:2001kd}
\begin{eqnarray}
\tilde{\alpha}=\cos(\theta_f-\theta_i)\cos\phi+i\cos(\theta_f+\theta_i) 
\sin \phi\,, \nonumber \\ 
\tilde{\beta}=\sin(\theta_f-\theta_i)\cos\phi-i \sin(\theta_f+\theta_i) 
\sin \phi\,,~
\label{ab2}
\end{eqnarray}
where 
\begin{equation}
\phi=\int_0^L\!\omega(x) dx \,,
\label{phi}
\end{equation}
and $\theta_i$ and $\theta_f$ are the values of the mixing angle in
matter $\hat{\theta}_{12}$ at the initial and final points of neutrino
evolution.  To a good accuracy, the matter density profile of the
Earth is spherically symmetric, so that one can set
$\theta_f=\theta_i$; however, for generality we shall allow for
different values of $\theta_i$ and $\theta_f$.

Substituting (\ref{ab2}) into Eq.~(\ref{P2e1}), one finds  
\begin{equation}
P_{2e}^\oplus-P_{2e}^{(0)}=c_{13}^2 
\left[
\sin(\theta_f-\theta_i) \sin(\theta_f-\theta_i+2\theta_{12}) \cos^2 \phi+ 
\sin(\theta_f+\theta_i)\sin(\theta_f+\theta_i-2\theta_{12}) \sin^2 \phi\right].
\label{P2e5}
\end{equation}
To leading order in $V_{i,f}\,$ this simplifies to  
\begin{equation}
P_{2e}^\oplus-P_{2e}^{(0)}=c_{13}^4\,\sin^2 2\theta_{12}\,\frac{1}{2}
\left[\frac{V_f-V_i}{2\delta}\cos^2 \delta L+\frac{V_i+V_f}{2\delta} 
\sin^2 \delta L \right] \,.
\label{P2e6}
\end{equation}

It is easy to make sure that for all the special cases discussed above
our formula (\ref{P2e7}) exactly reproduces the corresponding leading-order 
in $V$ expressions. The cases of one layer of constant density
and three layers of constant densities [Eqs.~(\ref{PND4}) and
(\ref{P2e4})] are immediately obtained from Eq.~(\ref{P2e7}) by a
straightforward calculation of the integral with the corresponding
matter density profiles.  The adiabatic case is recovered when one
integrates in Eq.~(\ref{P2e7}) by parts and neglects the term with the
derivative of the potential (i.e.  keeps only the off-integral term).
This gives
\begin{eqnarray}
P_{2e}^{\oplus}-P_{2e}^{(0)}\simeq 
c_{13}^4 \sin^2 2\theta_{12}\, 
\frac{1}{4\delta}\,\left\{V(r)\cos[2\delta\cdot(L-r)]\right\}
|_{0}^{L} \nonumber \\
=c_{13}^4 \sin^2 2\theta_{12}\, \frac{1}{4\delta}\,\left\{V_f-V_i 
\cos 2\delta L\right\}\,,
\label{P2e8}
\end{eqnarray}
which coincides with Eq.~(\ref{P2e6}). 

Quite analogously, in all these special cases Eq.~(\ref{eq:adiab}) exactly 
reproduces the corresponding leading-order in $V/2\delta$ expressions for 
$P_{2e}^\oplus-P_{2e}^{(0)}$.

\section{Comparison with numerical results}
\label{sec:comp-with-numer}

We now proceed to test the accuracy of our approximations by comparing
the results obtained using our main formulas in Eqs.~(\ref{PND2}) and 
(\ref{P2e7}), as well as Eq.~(\ref{eq:adiab}), with those following from 
the exact numerical integration of the three--neutrino evolution 
equation employed in Ref.~\cite{Maltoni:2003da}.
\FIGURE{
  \centering  
\includegraphics[height=6.5cm,width=.7\linewidth,clip]{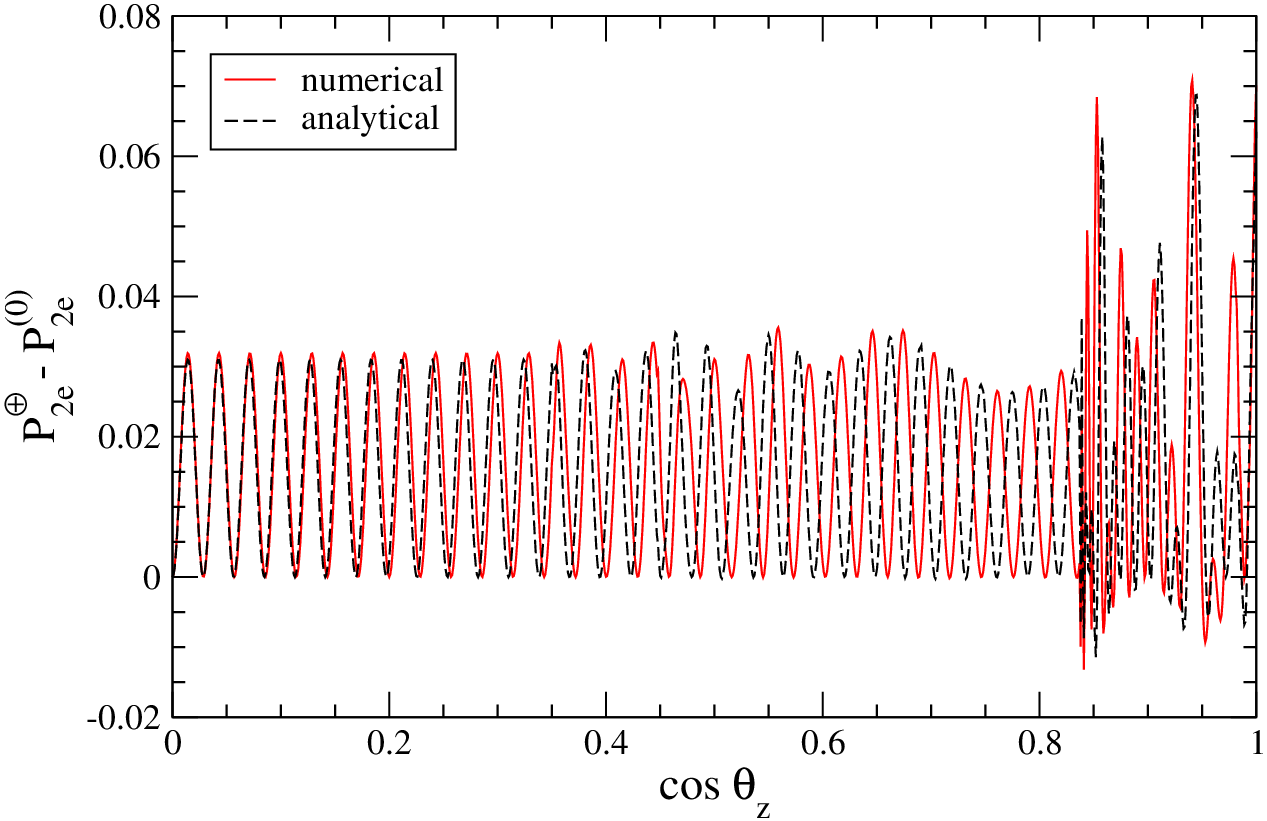}
  \caption{Two-flavour 
    Earth regeneration factor versus zenith angle for a 10 MeV solar
    neutrino from the exact numerical calculation and the analytic
    form given in Eq.~(\ref{P2e7}).}
  \label{fig:check-a}
  }

As has already been pointed out, for the validity of perturbation 
theory in $V$ the condition $V \ll 2\delta$ is not sufficient: one must also 
require that the correction to the oscillation phase, which is $\sim V L$, 
is small compared to unity, and not only compared to the main
contribution to the phase, $\delta L$. This is illustrated in
Fig~\ref{fig:check-a}, which gives the comparison of the predicted
probability differences due to the regeneration effect at the Earth,
$P_{2e}^\oplus-P_{2e}^{(0)}$, versus zenith angle, corresponding to
different solar neutrino arrival directions. The solid (red) curve
follows from the exact numerical calculation, while the dashed (black)
curve corresponds to the analytic formula in Eq.~(\ref{P2e7}).  One
sees clearly that the difference between the night-day probability
differences found in our analytic approximation and those that follow
from the exact numerical integration of the three--neutrino evolution
equation is maximal for neutrinos with small zenith angle.  For such
neutrinos with a longer path inside the Earth the difference between
the exact and approximate oscillation phases accumulates and becomes
relatively large, leading to significant differences between the exact
and analytic results, as can be seen from the figure.

\FIGURE{
  \centering
  \includegraphics[height=6.5cm,width=.7\linewidth,clip]{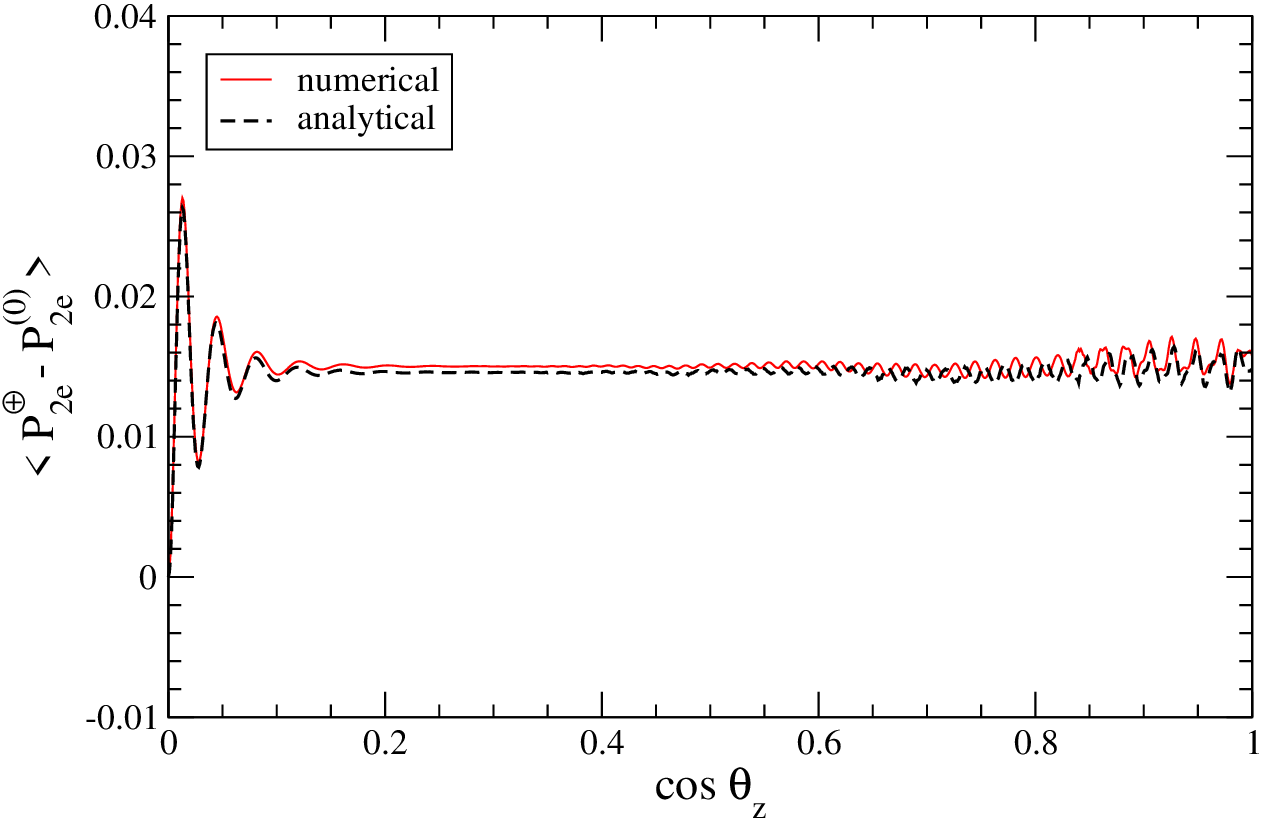}
  \caption{
    Two-flavour Earth regeneration factor versus zenith angle
    averaged for the Super-K detector, from the exact numerical
    calculation and the analytic formula in Eq.~(\ref{P2e7}).  }
  \label{fig:check-c}
  }

On the other hand, for large enough baselines, when $V L$ becomes of
order unity, also $\delta L\gg 1$, so that the averaging regime sets
in, either because of the integration over the baselines or, for a
fixed zenith angle, because of the integration over energies.  Note
that averaging over energies is always present in the experimental
data due to the fact that the neutrino energy is not directly measured
and also due to the finite energy resolution of the detectors.  In
addition to the energy resolution, for comparison with data one has
also to fold in the neutrino spectrum and relevant neutrino-detection
cross section.  All this leads to the smoothening of the above
difference. The net effect is that even order 1 shifts of the phase
become unimportant in practice, and neglecting the matter-induced
correction to the phase is justified. 
This is illustrated in Fig.~\ref{fig:check-c}, which compares the
probability differences $P_{2e}^\oplus-P_{2e}^{(0)}$ properly averaged
over the full energy range relevant for the Super-K detector.  It can
be seen from this figure that the averaging gives a much better
agreement between the analytic and exact results.
\FIGURE{
  \centering
  \includegraphics[height=6.5cm,width=.7\linewidth,clip]{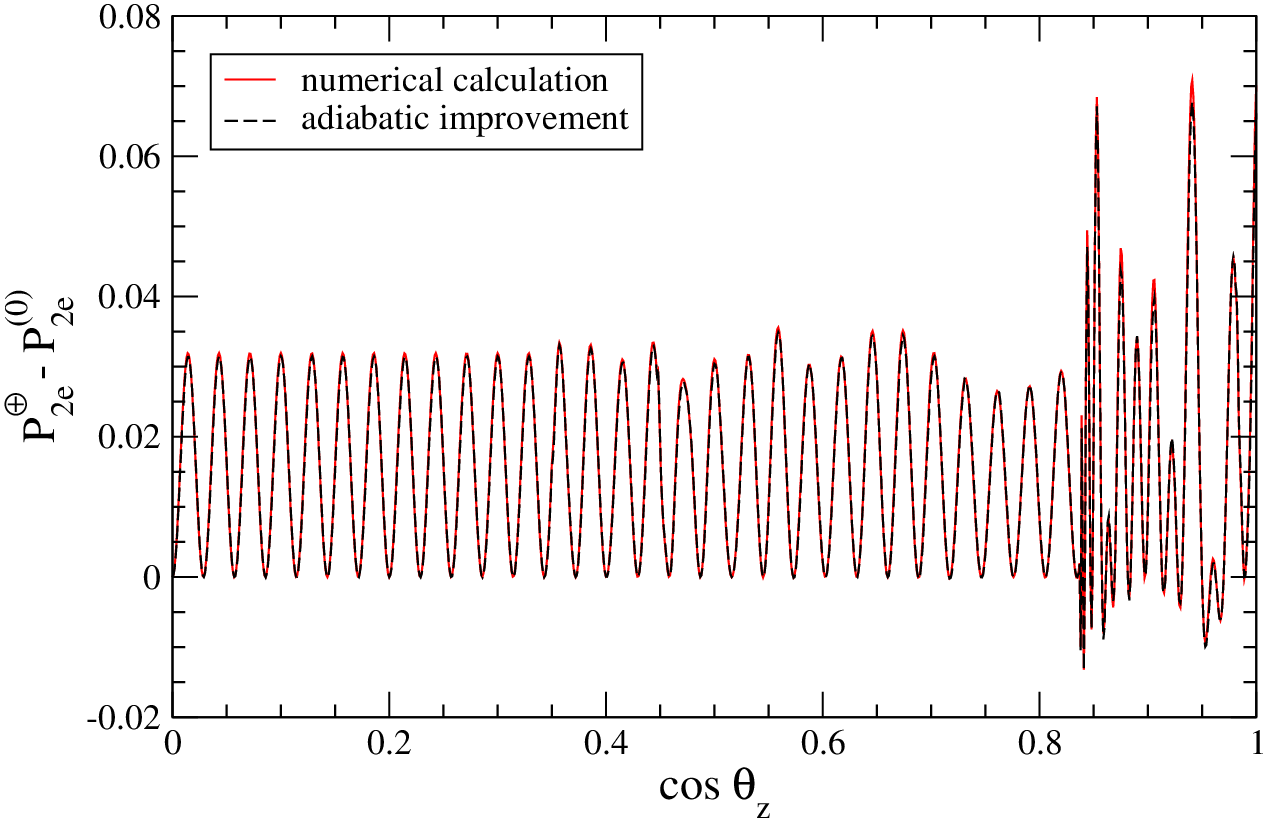}
  \caption{{Two-flavour 
      Earth regeneration factor versus zenith angle for a 10 MeV solar
      neutrino from the exact numerical calculation and the improved
      analytic formula given in Eq.~(\ref{eq:adiab})}.}
  \label{fig:check-b}
  }

Finally, note that due to relatively low statistics in the solar
neutrino experiments, integration over zenith angles is also usually
required. This also leads to an averaging over the oscillation phase,
giving a very good agreement between the analytic and exact numerical
results (see Fig.~\ref{fig:d-n-comp} below).

In the situations when no (or little) averaging over neutrino energies
or integration over zenith angles is involved, the improved 
perturbation theory formula given in Eq.~(\ref{eq:adiab}) yields much
better results.  This is illustrated by Fig.~\ref{fig:check-b} which
shows that in this case there is perfect agreement between the
analytic and numerical results.
\FIGURE{
  \centering
  \includegraphics[height=10cm,width=\linewidth,clip]{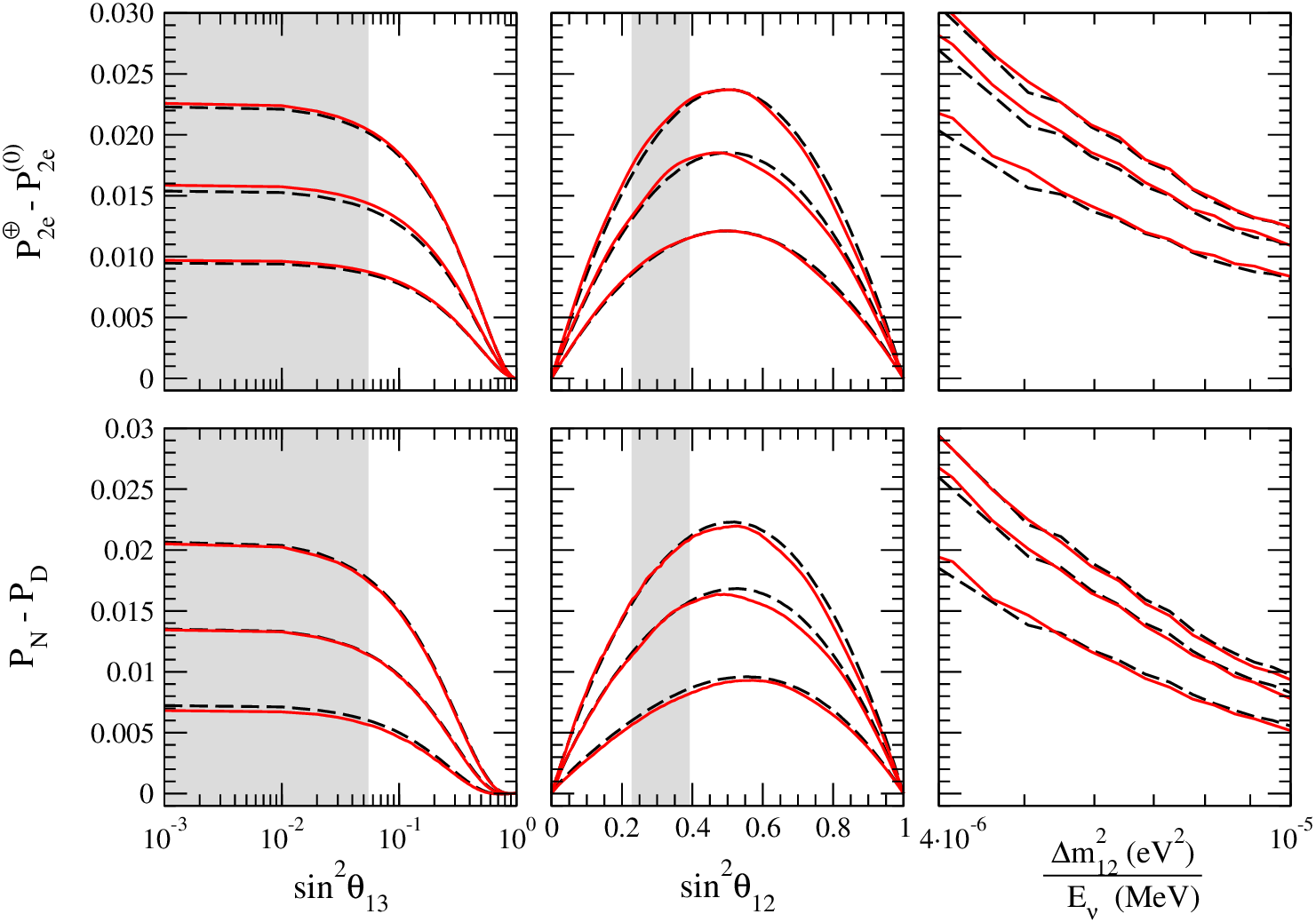}
  \caption{Comparison of predicted Earth regeneration factors and 
    day--night survival probability differences: solid lines
    correspond to exact numerical calculation, dashed ones are the
    results of the analytic formulas in Eqs.~(\ref{PND2}) and
    (\ref{P2e7}). See text for details.}
  \label{fig:d-n-comp}
  }

Finally, we turn to the case of interest of three-flavour day--night
solar flux differences.  Fig.~\ref{fig:d-n-comp} compares, for the
site of the Super-K detector, the analytic and numerical predictions
for the probability differences $P_{2e}^\oplus-P_{2e}^{(0)}$ and
$P_N-P_D$, integrated over the whole night and day ranges of the
zenith angles and averaged over one year. The results are given for
the neutrino energy $E=10$ MeV. 

The left panels give the dependence of these probability differences
on the angle $\theta_{13}$. 
The middle and right panels display the dependence on the ``solar''
mixing angle $\theta_{12}$ and solar mass--squared difference $\Delta
m_{21}^2$, respectively.
The vertical bands in the left and middle panels indicate the allowed
3$\sigma$ regions for $\sin^2 \theta_{13}$ and $\sin^2 \theta_{12}$
respectively.
The central curves in all panels are obtained for the best--fit values
of the undisplayed neutrino oscillation parameters, as obtained in
Ref.~\cite{Maltoni:2003da}. 
On the other hand the outer curves in all panels correspond to the
currently allowed $3\sigma$ range of the corresponding undisplayed
parameter. For example in the left panels the upper curves correspond
to $\sin^2\theta_{12} = 0.39$ and $\Delta m^2 = 5.4 \times 10^{-5}
~{\rm eV}^2$, while the lower ones correspond to $\sin^2\theta_{12} = 0.23$
and $\Delta m^2 = 9.5\times 10^{-5} ~{\rm eV}^2$, for energy $E = 10$ MeV.
In other words, the separation band between the upper and lowermost
curves in each panel gives a measure of the uncertainty in the
theoretical prediction allowed by current data.

Clearly, in all cases our approximate results reproduce very well the
results found from the exact numerical integration of the
three--neutrino evolution equation, with a precision much better than
current experimental sensitivities.

\section{Future experiments and sensitivity to $\theta_{13}$}
\label{sec:future}

The three-flavour day--night asymmetry in the solar neutrino flux
correlates with the magnitude of the leptonic mixing angle
$\theta_{13}$ ~\cite{Blennow:2003xw}. Adopting the standard definition
\begin{equation}
  \label{eq:Adn}
A_{ND} = 2 \frac{N-D}{N+D}
\end{equation}
and using (\ref{PND2}) and (\ref{P2e7}) one can see that the main
dependence on $\cos\theta_{13}$ is quadratic. Here we give a quantitative
study of this interesting correlation, as displayed in
Figs.~\ref{fig:Adn-theta13} and ~\ref{fig:Adn2-theta13}.
\FIGURE{
  \centering
 \includegraphics[height=7cm,width=.9\linewidth,clip]{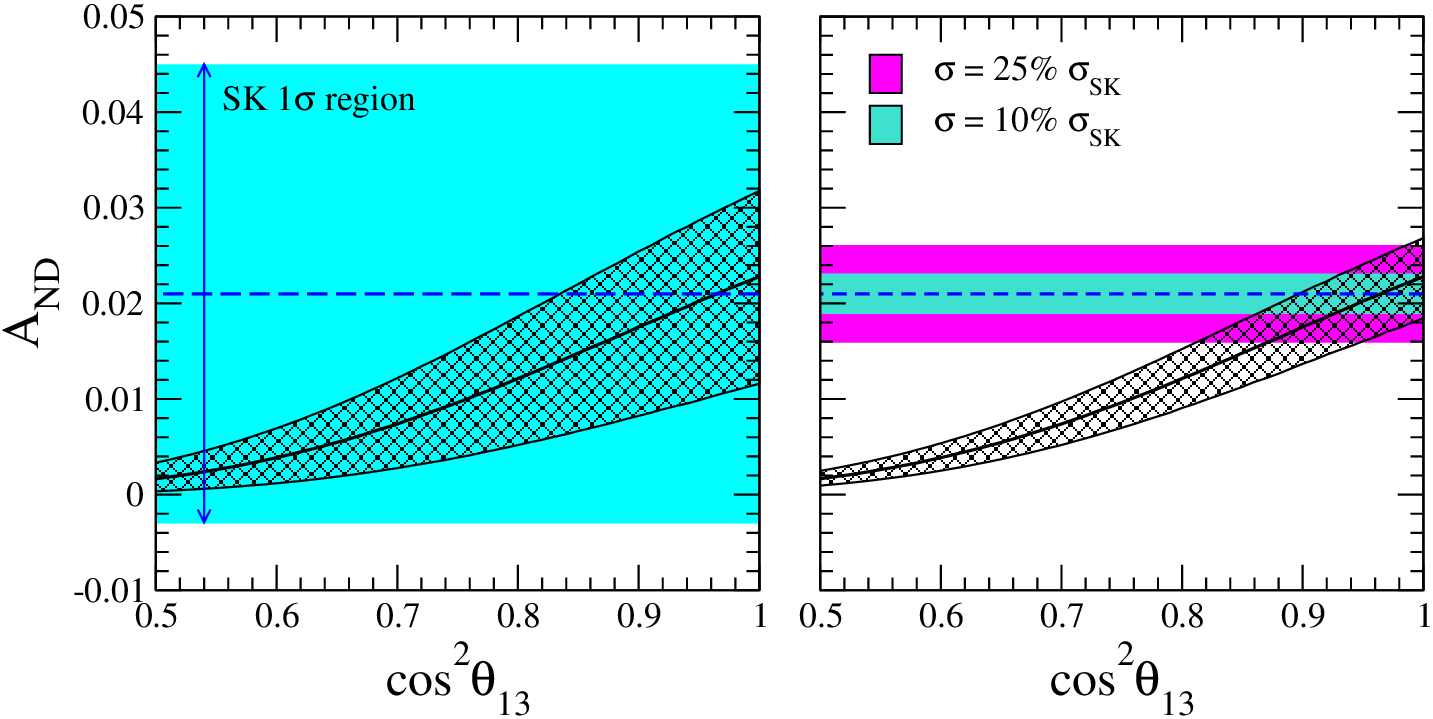}
 \caption{Day--night asymmetries in three-flavour
   solar neutrino oscillations versus the mixing angle $\theta_{13}$.
   The left panel shows the present status, while the right one gives
   the projected sensitivities in future water Cherenkov experiments
   assuming the same central value of $A_{ND}$.}
  \label{fig:Adn-theta13}
  }
The left panel in Fig.~\ref{fig:Adn-theta13} gives the magnitude of the
day-night asymmetry measured at the Super-K experiment $2.1 \% \pm 2.0
\% (\rm{stat}) \pm 1.3 \% (\rm{syst})$~\cite{Smy:2003jf} compared with
the range for this quantity which is theoretically predicted on the
basis of current analysis of solar neutrino data~\cite{Maltoni:2003da}.  
The horizontal band corresponds to the current $1\sigma$ uncertainty in 
the measured day-night asymmetry, while the hatched region shows the 
$3\sigma$ uncertainty implied by the current global determination of 
oscillation parameters. Clearly, the large errors with which the small 
day--night asymmetries are determined by current experiments are such that 
the asymmetry measurement does not play a significant role in constraining 
the angle $\theta_{13}$, the upper bound on which is dominated by other 
data in the global fit, mainly by the reactor data. However, the simple 
and direct correlation between the predicted asymmetries and $\theta_{13}$
that can be appreciated in Figs.~\ref{fig:Adn-theta13} and 
\ref{fig:Adn2-theta13} can be used to provide an estimate of 
the degree to which the mixing angle $\theta_{13}$ can be probed at a
future generation of water Cherenkov solar neutrino experiments. The 
significance of the future data for the determination of $\theta_{13}$ 
will depend substantially on both the central value and the errors of the 
measured $A_{ND}$.

\FIGURE{
  \centering
 \includegraphics[height=7cm,width=.9\linewidth,clip]{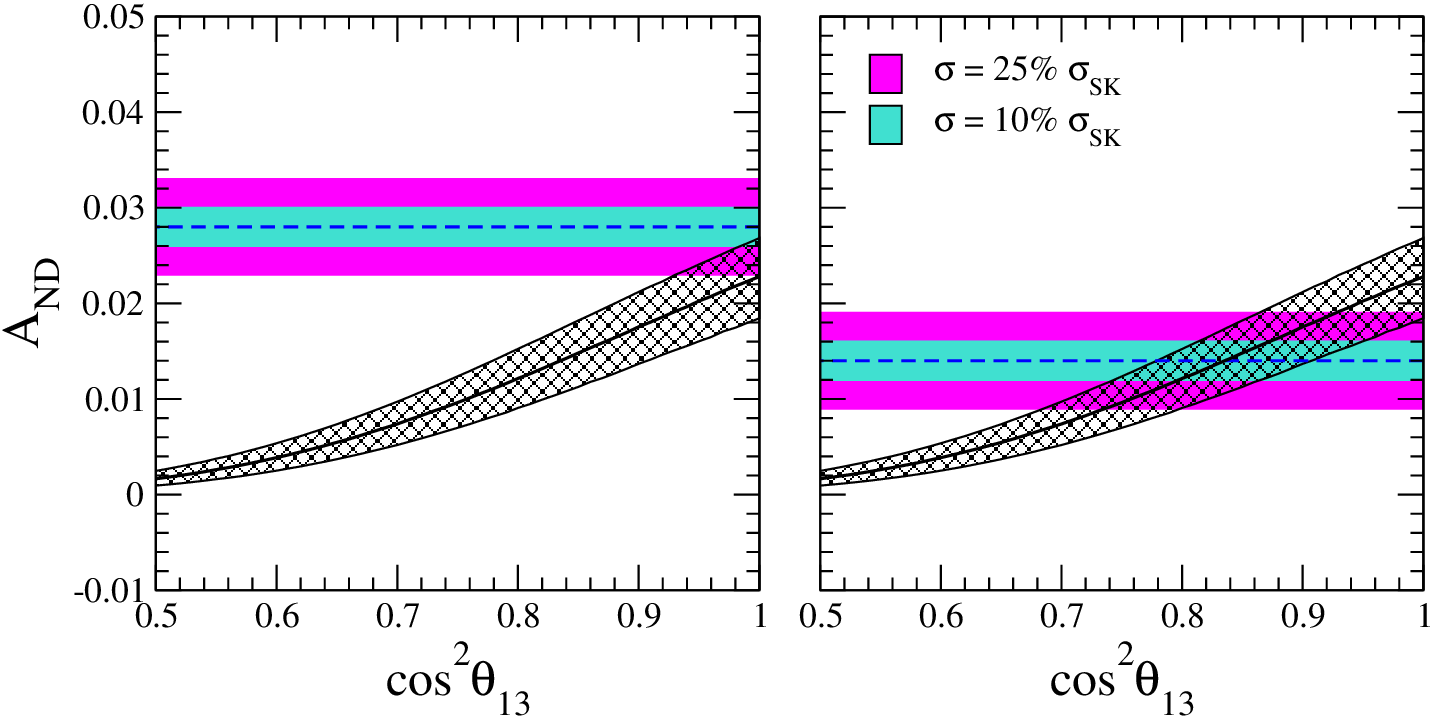}
 \caption{Same as the right panel of Fig.~\ref{fig:Adn-theta13}, 
 but for different central values of expected measured $A_{ND}$: 
 2.8\% (left panel) and 1.4\% (right panel). }
  \label{fig:Adn2-theta13}
  }

The horizontal bands in the right panel of Fig.~\ref{fig:Adn-theta13}
and in both panels of Fig.~\ref{fig:Adn2-theta13} provide an estimate
of the expected improved day-night asymmetry measurements at a larger
Super-K--like water Cherenkov detector, with $1\sigma$ errors at 25\%
and 10\% of the current Super-K error. Such errors might be realistic at 
future detectors. For example, the Underground Nucleon decay and Neutrino 
Observatory (UNO) experiment is expected to improve the sensitivity on the 
day-night asymmetry down to $\sigma=0.25\,\sigma_{SK}$, while for the 
Hyper-Kamiokande experiment the errors can be much smaller~\cite{UNO}.

On the other hand, the hatched regions in the right panel of
Fig.~\ref{fig:Adn-theta13} and in Fig.~\ref{fig:Adn2-theta13}
correspond to the $3\sigma$ range for solar neutrino oscillation
parameters expected after improved KamLAND results. Although further
improvements on the determination of solar neutrino oscillation
parameters are likely to be available by then, the figure shows how
the uncertainty in the day--night asymmetry measurement dominates the
projected accuracy on the mixing angle $\theta_{13}$.

Let us first assume that the central value of $A_{ND}$ measured in
future experiments coincides with the current value found by Super-K
(right panel of Fig.~\ref{fig:Adn-theta13}). As one can see from the
figure, no improvement on the current upper bound on $\theta_{13}$ can
be obtained in that case.  However, it is quite possible that the
future central value of the measured night-day asymmetry will differ
from the currently measured one, at least within the current $1\sigma$
experimental error \footnote{Future best-fit values of the parameters
  $\theta_{12}$ and $\Delta m_{12}^2$, which determine the position of
  the hatched regions in Figs.~\ref{fig:Adn-theta13} and
  ~\ref{fig:Adn2-theta13}, can also differ from the current ones.
  However, the effect of this deviation cannot be as large as that for
  $A_{ND}$ since their present-day errors are significantly smaller.}.
This situation is illustrated in Fig.~\ref{fig:Adn2-theta13}. As one
can see from the left panel, if the future central value of measured
$A_{ND}$ is higher than the present Super-K one, the current upper
limit on $\theta_{13}$ can actually be improved. If, on the contrary,
a lower value of $A_{ND}$ is measured (right panel of
Fig.~\ref{fig:Adn2-theta13}), the derived upper limit on $\theta_{13}$
will be substantially weaker than the current one and thus irrelevant.
However, as can be seen from the figure, in that case a {\em lower}
bound on $\theta_{13}$ may appear; together with the current upper
bound it may actually lead to a rather precise determination of
$\theta_{13}$.

Thus, an improved measurement of the day--night asymmetry can be immediately 
converted into more precise information on $\theta_{13}$.  

It should be noted that, even though the day-time solar neutrino signal 
depends on $\theta_{13}$ more sensitively than the day-night asymmetry 
(essentially as $c_{13}^4$), this dependence is always multiplied by the 
overall normalization of the solar neutrino flux. For $^8\!B$ neutrinos, 
the uncertainty in the flux normalization factor $f_B$ is of the same size as 
the possible effect of non-zero $\theta_{13}$. This makes it difficult to 
disentangle the two effects and hinders the precise determination of both 
$f_B$ and $\theta_{13}$ from the charged-current data.  In contrast to this, 
the day-night asymmetry is independent of the overall flux normalization and 
so may provide an unambiguous information on $\theta_{13}$.

\section{Discussion and outlook}
\label{sec:disc-outlook}

We have derived a simple and accurate analytic expression for the
day-night difference of the flux of solar $\nu_e$ coming to
terrestrial detectors, with 3-flavour effects fully taken into
account. Our approach was based on a simple perturbation theory in the
matter-induced potential of neutrinos inside the Earth, without any
assumptions regarding the Earth's density profile. Our results are
therefore valid for an arbitrary density profile.

We have checked our analytic formula by comparing it with the results
of the exact numerical calculations with the 3-flavour evolution
equation, and found that the accuracy of the analytic approach is
typically about a few per cent when the neutrino path length inside
the Earth is small, or when the integration over the zenith angles or
averaging over neutrino energies are performed.
On the other hand, when both the neutrino energy and zenith angle are
fixed, the formula in Eq.~(\ref{P2e7}) does not provide a sufficient 
accuracy. For this case we have derived an improved perturbation-theoretic
formula [Eq.~(\ref{eq:adiab})]. Compared to the expression in 
Eq.~(\ref{P2e7}) it contains the adiabatic oscillation phase in the 
integrand instead of the in-vacuum one. The improved formula gives a
perfect agreement with the exact numerical results even when no
averaging over the zenith angles or neutrino energy (i.e., over the
oscillation phase) is performed.

We have studied the dependence of our results on the leptonic mixing
angle $\theta_{13}$ which is of great interest for many reasons, most
importantly because it governs CP violation in neutrino oscillations.
We have found that for an arbitrary Earth density profile the
day-night difference of the solar neutrino flux at the detector scales
mainly as $c_{13}^6$, as previously found for the constant density
profile in \cite{Blennow:2003xw}.  The remaining (milder) dependence
stems primarily from the $\theta_{13}$-dependence of the neutrino
mixing angle in matter at the production point inside the Sun.

Although the smallness of $\theta_{13}$ implies that the solar
neutrino data, including the day-night asymmetry, should depend rather
weakly on this parameter, this dependence may not be negligible. For
example, the current $3\sigma$ limit $\sin^2\theta_{13}<0.054$
\cite{Maltoni:2003da} means that the day-night difference of the solar
neutrino signal can be suppressed by up to 15\% compared to the
2-flavour (i.e. $\theta_{13}=0$) case. If one fixes $\Delta
m_{31}^2=2\times 10^{-3}$ eV$^2$ (which is the best-fit value coming
from the analysis of the atmospheric neutrino oscillations performed
by the \mbox{Super-K} collaboration~\cite{SKatm04}), then the
$3\sigma$ upper bound is $\sin^2\theta_{13}<0.066$
\cite{Maltoni:2003da}, and the day-night difference of the solar
neutrino signal can be suppressed by as much as 20\% compared to the
2-flavour case.

While the accuracy with which the solar neutrino day-night effect is
measured by the current experiments is insufficient for probing the
value of $\theta_{13}$, future very large water Cherenkov detectors,
such as UNO or Hyper-Kamiokande, may be able to provide a significant
information on it.
Depending on the measured value of $A_{ND}$, the current upper bound
on $\theta_{13}$ may be improved, or even a lower bound on this mixing
parameter may be obtained. 
An advantage of the day-night asymmetry as opposed to the day-time  
signal is that $A_{ND}$ is independent of the overall normalization of
the solar neutrino flux, which is currently not known with sufficient
accuracy.

It should be noted that the Earth matter effects on the oscillations
of the supernova neutrinos inside the Earth are governed by the same
quantity, $P_{2e}^\oplus-P_{2e}^{(0)}$, which determines the Earth
matter effects on the solar neutrino oscillations (see, e.g.,
\cite{Dighe:1999bi}).  Therefore our results, Eqs.~(\ref{P2e7}) and
(\ref{eq:adiab}), can also be used for studying the supernova neutrino
oscillations inside the Earth.  Note, however, that the typical
energies of supernova neutrinos $(E\sim 30$ MeV) are higher than those
of solar neutrinos, and so the expected accuracy of our approximation
for supernova neutrinos is $\sim$ (10 -- 15)\%.

\acknowledgments

This work was supported by Spanish grant BFM2002-00345, by the
European Commission RTN network HPRN-CT-2000-00148 and by the European
Science Foundation network grant N.~86. E.~A. is supported by a
sabbatical grant N. SAB2002-0069 of the Spanish MECD. M.A.T.\ is
supported by the M.E.C.D.\ fellowship AP2000-1953.

\vspace*{4mm} {\em Note added.} As we completed our paper, two new
papers appeared \cite{deHolanda:2004fd,Ioannisian:2004jk}, in which
analytic formulas for the Earth matter effect on neutrino
oscillations, similar to ours, were derived. The
first~\cite{deHolanda:2004fd} is based on adiabatic perturbation
theory, while Ref.~\cite{Ioannisian:2004jk} employs ordinary
perturbation theory, similar to that used in the present paper.  Note,
however, that Refs.~\cite{deHolanda:2004fd,Ioannisian:2004jk} adopt
the two-flavour approximation, while our analysis is performed in the
three-flavour framework.

\begin{appendix}
\label{app:derivation}
\renewcommand{\theequation}{\thesection\arabic{equation}}
\setcounter{equation}{0}
\section{Derivation of Eq.~(\ref{eq:adiab})}

We shall derive here expression (\ref{eq:adiab}) for the Earth regeneration 
factor $P_{2e}^{\oplus}-P_{2e}^{(0)}$. In contrast to perturbation theory 
in $V$ developed in sec.~\ref{sec:an-analyt-expr}, we employ here perturbation 
theory in the small parameter $V/2\delta$, so that the obtained results will 
be valid for arbitrary values of $V L$. 

Consider the neutrino evolution equation in the mass-eigenstate basis, 
$i(d/dt)\nu_{mass}=H_{mass}\nu_{mass}$. Here $\nu_{mass}=U^\dag \nu$, 
and the Hamiltonian $H_{mass}=U^\dag H U$ can be written explicitly as   
\begin{equation}
H_{mass}=\left(\begin{array}{lll}
c_{12}^2\,c_{13}^2\,V & ~~~s_{12}\,c_{12}\,c_{13}^2 V & ~~~c_{12}\,s_{13}\,
c_{13}\,e^{-i\delta_{\rm CP}}\,V \\ 
s_{12}\,c_{12}\,c_{13}^2\,V & ~~~s_{12}^2\,c_{13}^2\,V+2\delta & ~~~s_{12}\,
s_{13}\,c_{13}\,e^{-i\delta_{\rm CP}}\,V \\
c_{12}\,s_{13}\,c_{13}\,e^{i\delta_{\rm CP}}\,V & ~~~s_{12}\,s_{13}\,c_{13}\,
e^{i\delta_{\rm CP}}\,V & ~~~2\Delta + s_{13}^2\,V \end{array}\right).
\label{A1}
\end{equation}
Since $V\ll 2\delta \ll 2\Delta$ and $s_{13}\ll 1$, one can neglect the 
(1-3) and (2-3) elements of $H_{mass}$ compared to the (3-3) element. This 
means that the evolution of the third mass eigenstate $\nu_3$ essentially 
decouples from that of the first two eigenstates and, in addition, that matter 
effects on the third mass eigenstate are negligible. 

Diagonalization of the $(\nu_1,\,\nu_2)$-subsector of $H_{mass}$ is performed 
through the rotation by the angle $\theta'(t)$ which is determined 
according to  
\begin{equation}
\cos 2\theta'(t)=\frac{\delta-\cos 2\theta_{12}\,c_{13}^2 V(t)/2}
{\omega(t)}\,,\qquad~~
\sin 2\theta'(t)=\frac{\sin 2\theta_{12}\,c_{13}^2 V(t)/2}
{\omega(t)}\,,
\label{theta'}
\end{equation}
with $\omega(t)$ given in Eq.~(\ref{omega}). Note that $\theta'=
\hat{\theta}_{12}-\theta_{12}$, where $\hat{\theta}_{12}$ was defined in 
Eq.~(\ref{cos}). The smallness of the potential $V$ compared to $2\delta$ 
means that the neutrino matter eigenstates almost coincide with 
mass eigenstates, i.e. $\theta'(t)\ll 1$. We shall now employ perturbation 
theory in $\sin\theta'$, which is essentially the same as perturbation 
theory in $V/2\delta$. 

The decoupling of $\nu_3$ allows one to write 
\begin{equation} 
H_{mass}\simeq\omega(t) \left(\begin{array}{ccc}
-\cos 2\theta'(t) & \sin 2\theta'(t) & 0 \\ 
\sin 2\theta'(t) & \cos 2\theta'(t) & 0 \\
0 & 0 & 2\tilde{\Delta}/\omega(t)\end{array}\right)\,.
\label{A3}
\end{equation} 
Here $\tilde{\Delta}=\Delta-(V/2+\delta)/2$, and we have redefined the 
common phase of the neutrino states so as to make the Hamiltonian of the 
$(\nu_1,\,\nu_2)$ sector traceless and used Eq.~(\ref{theta'}). The 
Hamiltonian $H_{mass}$ in Eq.~(\ref{A3}) can be decomposed into the term of 
zeroth order in $s'\equiv \sin\theta'$ and term that contains first and higher 
orders in $s'$ according to 
\begin{equation} 
H_{mass}\simeq \left(\begin{array}{ccc}
-\omega(t) & 0 & 0 \\ 0 & \omega(t) & 0 \\0 & 0 & 2\tilde{\Delta}
\end{array}\right)
+2 \omega(t)\left(\begin{array}{ccc}
s'(t)^2 & s'(t) c'(t) & ~~0 \\ 
s'(t) c'(t)  & -s'(t)^2  & ~~0 \\
0 & 0 & ~~0 \end{array}\right)\equiv H_{m0}+H_{m1}\,. 
\label{A4}
\end{equation} 
To leading order in $s'(t)$, only the (1-2) and (2-1) elements of $H_{m1}$ 
are different from zero. The evolution matrix to first order in $s'$ can be 
found from a formula which coincides with Eq.~ (\ref{Stilde}) 
with $\tilde{S}_0$ and $\tilde{H}_1$ replaced by $(S_{mass})_0$ and 
$H_{m1}$, respectively. 
The zeroth-order evolution matrix in the mass eigenstates basis is 
$(S_{mass})_0={\rm diag}(e^{i\phi},\,e^{-i\phi},\,\tilde{f})$, and to 
first order in $s'$ we find
\begin{equation} 
S_{mass}\simeq \left(\begin{array}{ccc}
e^{i\phi} & -i C e^{i\phi} & ~~0 \\ -i C^* e^{-i\phi}  & e^{-i\phi}  & ~~0 \\ 
0 & 0 & ~~\tilde{f} \end{array}\right)\,,
\label{A5}
\end{equation} 
where 
\begin{equation}
\phi=\int_{t_0}^{t}\omega(t') dt'\,,\qquad\quad
C=2 \int_{t_0}^{t}\omega(t') s(t') e^{-2i\int_{t_0}^{t'}\omega(x) dx} dt'\,.
\label{A6}
\end{equation}
The parameter $\tilde{f}$ is not relevant to our discussion. The amplitude 
$A^\oplus(\nu_2\to \nu_e)$ is obtained as 
\begin{equation}
A^\oplus(\nu_2\to \nu_e)=(1 \quad 0\quad 0)\; U S_{mass} 
\left(\begin{array}{c} 
0 \\ 1 \\ 0
\end{array}\right)=c_{13}(-ic_{12} C e^{i\phi} + s_{12} e^{-i\phi})\,. 
\label{A7}
\end{equation}
This gives
\begin{equation}
P_{2e}^{\oplus}-P_{2e}^{(0)}=c_{13}^2 \sin 2\theta_{12}\,{\rm Im}(e^{2i\phi} 
C)\,,
\label{A}
\end{equation}
which immediately leads to Eq.~(\ref{eq:adiab}). 

The two-flavour limit of Eq.~(\ref{eq:adiab}) was derived in 
Ref.~\cite{Ioannisian:2004jk} by dividing the interval $[0,\,L]$ into small 
intervals of constant matter densities.  

\end{appendix}

\end {document}